\documentclass[11pt]{article}
\usepackage{graphicx}
\usepackage{amsmath,amssymb}
\def\1{{\bf 1}}

\def\be{\begin{equation}}
\def\ee{\end{equation}}

\begin{document}
\noindent {\bf Amyloids: Composition, Functions and Pathology, NOVA Publishers (https://www.novapublishers.com), 2011, accepted.}\\
\noindent \rule{\textwidth}{1pt}\\

\bigskip

\noindent {\Large\bf Atomic-resolution structures of prion AGAAAAGA amyloid fibrils}

\bigskip

\noindent {\Large Jiapu Zhang}\\
School of Sciences, Information Technology and Engineering,
University of Ballarat, Mount Helen, Ballarat, Victoria 3353, Australia,
Phone: 61-423487360, 61-3-5327 9809, Email: jiapu\_zhang@hotmail.com, j.zhang@ballarat.edu.au

\bigskip

\noindent {\bf Abstract:} To the best of the author's knowledge, there is little structural data available on the AGAAAAGA palindrome in the hydrophobic region (113-120) of prion proteins due to the unstable, noncrystalline and insoluble nature of the amyloid fibril, although many experimental studies have shown that this region has amyloid fibril forming properties and plays an important role in prion diseases. In view of this, the present study is devoted to address this problem from computational approaches such as local optimization steepest descent, conjugate gradient, discrete gradient and Newton methods, global optimization simulated annealing and genetic algorithms, canonical dual optimization theory, and structural bioinformatics. The optimal atomic-resolution structures of prion AGAAAAGA amyloid fibils reported in this Chapter have a value to the scientific community in its drive to find treatments for prion diseases or at least be useful for the goals of medicinal chemistry.\\


\section{INTRODUCTION}
Prion diseases are invariably fatal and highly infectious neurodegenerative diseases that affect humans and animals. Prion diseases are amyloid fibril diseases. Prion amyloid fibrils are believed rich in $\beta$-sheet structure and contain a cross-$\beta$ core. Many experimental works \cite{brown2000,brown2001,brown1994,holscher1998,jobling2001,jobling1999,kuwata2005,norstrom2005,wegner2002} show that the hydrophobic region AGAAAAGA of prion proteins (113-120) plays an important role in the conversion of amyloid fibrils. PrP lacking / deleting the palindrome (PrP 113-120) neither converted to PrP$^\text{Sc}$(infectious prion) nor generated proteinase K-resistant PrP \cite{brown2001,holscher1998,norstrom2005,wegner2002}. Brown et al. \cite{brown2000,brown1994} pointed out that the AGAAAAGA peptide was found to be necessary (though not sufficient) for blocking the toxicity and amyloidogenicity of PrP 106-126. The peptide AGAA did not form fibrils but the peptide AGAAAAGA formed fibrils in both water and PBS \cite{brown2000}. Thus, the minimum sequence necessary for fibril formation should be AGAAA, GAAAA, AAAAG, AAAGA, AGAAAA, GAAAAG, AAAAGA, AGAAAAG, GAAAAGA or AGAAAAGA. According to Brown \cite{brown2000}, AGAAAAGA is important for fibril formation and is an inhibitor of PrP$^\text{Sc}$ neurotoxicity.\\

\begin{figure}[h!]
\centerline{
\includegraphics[scale=0.5]{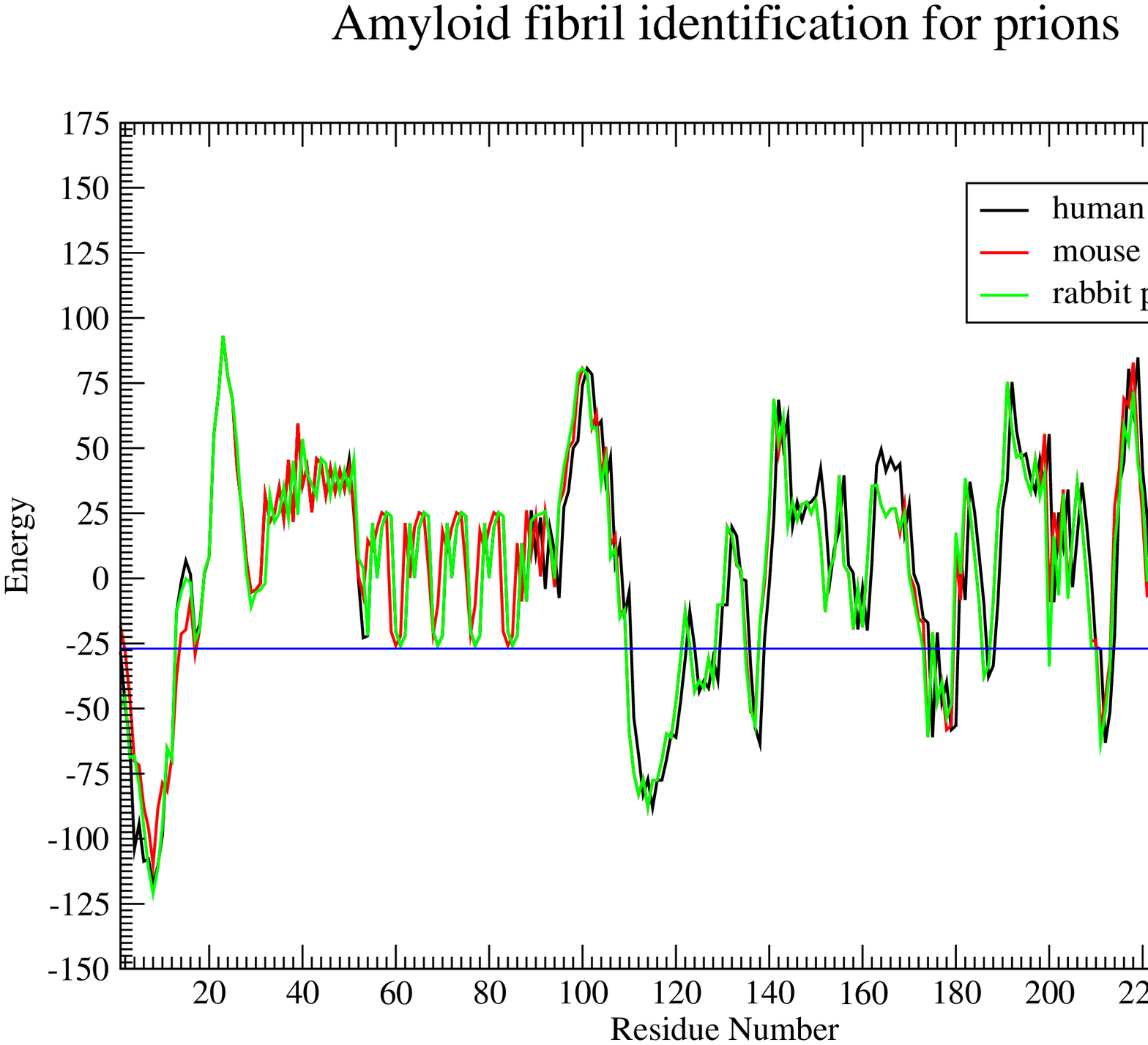}
}
\caption{Prion AGAAAAGA (113–-120) is clearly identified as fibril formation segment.}
\label{identification}
\end{figure}

In theory, for the sake of clarity, we use a program in Zhang et al. \cite{zhang2007} to confirm that prion AGAAAAGA (113-120) segment has an amyloid fibril forming property. The theoretical computation results are shown in Fig. \ref{identification}, from which we can see that prion AGAAAAGA (113-120) region is clearly identified as the amyloid fibril formation region because the energy is less than the amyloid fibril formation threshold energy -26 \cite{zhang2007}.
\\ 

However, due to the unstable, noncrystalline and insoluble nature of the amyloid fibril, to date structural information on AGAAAAGA region (113-120) has been very limited. This region falls just within the N-terminal unstructured region PrP (1-123) of prion proteins. Traditional X-ray crystallography and nuclear magnetic resonance (NMR) spectroscopy experimental methods cannot be used to get its structural information. Under this background, computational approaches or introducing novel mathematical formulations and physical concepts into molecular biology can significantly stimulate the development of biological and medical science. The author has introduced novel mathematical global and local optimization computational approaches to produce the atomic-resolution structures of prion (113-120) AGAAAAGA amyloid fibrils, which are in Section 2. Section 3 summarizes this article.\\

\section{COMPUTATIONAL OPTIMIZATION APPROACHES AND THEIR AGAAAAGA STRUCTURAL RESULTS}
In 2007, Sawaya et al. got a breakthrough finding: the atomic structures of all amyloid fibrils revealed steric zippers, with strong vdw interactions (LJ) between $\beta$-sheets and hydrogen bonds (HBs) to maintain the $\beta$-strands \cite{sawaya2007}. In this section, we will use suitable templates 2OMP.pdb (the LYQLEN peptide derived from human insulin residues 13–18), 1YJP.pdb (the GNNQQNY peptide from the yeast prion protein Sup35), 3FVA.pdb (NNQNTF 173-178 segment from elk prion protein), 3NHC.pdb (GYMLGS segment 127-132 from human prion with M129), 3NVE.pdb (MMHFGN segment 138-143 from Syrian Hamster prion), 3NVF.pdb (IIHFGS segment 138-143 from human prion), 3NVG.pdb (MIHFGN segment 137-142 from mouse prion) and 3NVH.pdb (MIHFGND segment 137-143 from mouse prion) from the Protein Data Bank (http://www.rcsb.org/) to construct some amyloid fibril models for the prion AGAAAAGA region (113-120).

\subsection{2OMP, 1YJP MODELS}
The author used the unmerge, mutate, and merge modules of Insight II (http://accelrys.com/) for 2OMP.pdb, 1YJP.pdb to build the 12-chain AGAAAA 2OMP-Model (MODEL01), 10-chain AGAAAAG 1YJP-Model (MODEL02), and 10-chain GAAAAGA 1YJP-Model (MODEL03) \cite{zhang2011a}. Then, using AMBER 10 \cite{case2008}, the  MODEL01-03 were refined by the hybrid of steepest descent (SD) and conjugate gradient (CG) -- simulated annealing (SA) -- SDCG methods, where SA phase made the MODEL01-03 reach sufficient equilibration and stability. MODEL1 (Fig. \ref{MODELS01-03}) belongs to models of Class 7 ($\beta$-strand antiparallel, face=back, up–up) of \cite{sawaya2007}, and MODEL02-03 (Fig. \ref{MODELS01-03}) belong to models of Class 1 ($\beta$-strand parallel, face-to-face, up–up) of \cite{sawaya2007}.\\

Sawaya et al. \cite{sawaya2007} proposed eight classes of steric zipper structures for peptide segments of fibril forming proteins. For each Class, Zhang \cite{zhang2011a} constructed the molecular structures for the hydrophobic region AGAAAAGA palindrome of prion proteins (113–120). Besides the above successful MODEL01-03, the unsuccessful molecular modeling experiences are: (1) For Class 1, based on the NNQQNY peptide from yeast prion protein Sup35 (1YJO.pdb), a hexamer model with six AAAAGA chains can be constructed by SDCG but cannot pass SA; (2) For Class 2 (i.e. $\beta$-strand parallel, face-to-back, up–-up), a tetramer model with four AGAAAA chains can be constructed basing on the SNQNNF peptide of human prion 170–175 (2OL9.pdb) by SDCG but cannot pass SA; (3) For other Classes (yeast Sup35 GNNQQNY form 2, 2OMM.pdb; yeast Sup35 NNQQ form 1, 2ONX.pdb; yeast Sup35 NNQQ form 2, 2OLX.pdb; human insulin VEALYL, 2OMQ.pdb; human Tau protein VQIVYK, 2ON9.pdb; human A$\beta$ GGVVIA, 2ONV.pdb; human A$\beta$ MVGGVV form 1, 2ONA.pdb; human A$\beta$ MVGGVV form 2, 2OKZ.pdb; bovine RNase SSTSAA, 2ONW.pdb) in Sawaya et al. \cite{sawaya2007}, the $\beta$-sheet structure of prion AGAAAAGA palindrome cannot pass the SDCG phase.

\subsection{3FVA MODELS}
Instead of using Insight II, Zhang et al. used the mutate module of the free package Swiss-PdbViewer (SPDBV Version 4.01) (http://spdbv.vital-it.ch) and the hybrid discrete gradient (DG) simulated annealing method (i.e. DGSA) to build the prion AGAAAAGA amyloid fibril models \cite{zhang2011b}. The models were built based on the template 3FVA.pdb of NNQNTF segment from elk prion protein (173-178). After the models were built, the refinements were done completely same as \cite{zhang2011a}. A six chains AGAAAA model could not successfully pass SA of Amber 10. However, two prion AGAAAAGA palindrome amyloid fibril models (Fig. \ref{MODELS04-05}) - a six chains GAAAAG model (MODEL04) and a six chains AAAAGA model (MODEL05) - were successfully obtained. Compared with \cite{zhang2011a}, the variations of RMSD, PRESS, and VOLUME (DENSITY) of the 4,400 ps' equilibration at 100 K are larger \cite{zhang2011b}; this might imply that DGSA is a little worse than Insight II.

\subsection{3NHC, 3NVE/F/G/H MODELS}
Replacing the DGSA of Subsection 2.2, in this Subsection we will use any optimization solver, which can accurately solve an optimization problem with 3 or 6 variables, to build the models; thus the methods in this subsection are very simple and general for any problems in molecular modeling research area. The model building templates are: 3NHC.pdb (GYMLGS segment 127-132 from human prion with M129), 3NVE.pdb (MMHFGN segment 138-143 from Syrian Hamster prion), 3NVF.pdb (IIHFGS segment 138-143 from human prion), 3NVG.pdb (MIHFGN segment 137-142 from mouse prion), and 3NVH.pdb (MIHFGND segment 137-143 from mouse prion). The mathematical theory is described as follows.\\

The atomic structures of all amyloid fibrils revealed steric zippers, with strong vdw interactions (LJ) between $\beta$-sheets and hydrogen bonds (HBs) to maintain the $\beta$-strands \cite{sawaya2007}. In mathematics, the potential energy mathematical formula for the vdw interactions between $\beta$-sheets is
\begin{equation} \label{LJ_AB_form}
V_{LJ}(r)=\frac{A}{r^{12}} -\frac{B}{r^6},
\end{equation}
and the potential energy mathematical formula for the HBs between the $\beta$-strands is 
\begin{equation} \label{HB_r_form}
V_{HB}(r)= \frac{C}{r^{12}} -\frac{D}{r^{10}} ,
\end{equation}
where $A,B,C,D$ are constants given. When $V_{LJ}$ and $V_{HB}$ are reaching their minimal values, the amyloid fibril structures should be in a most stable state. This is a molecular distance geometry problem (MDGP) \cite{grosso2009}, which arises in the interpretation of NMR data and in the determination of protein structure [as an example to understand MDGP, the problem of locating sensors in telecommunication networks is a DGP. In such a case, the positions of some sensors are known (which are called anchors) and some of the distances between sensors (which can be anchors or not) are known: the DGP is to locate the positions of all the sensors. Here we look sensors as atoms and their telecommunication network as a molecule]. The three dimensional structure of a molecule with $n$ atoms can be described by specifying the 3-Dimensional coordinate positions $x_1, x_2, \dots, x_n \in \mathbb{R}^3$ of all its atoms. Given bond lengths $d_{ij}$ between a subset $S$ of the atom pairs, the determination of the molecular structure is
\begin{eqnarray}
(\mathcal{P}_0 ) \quad to \quad find \quad &x_1,x_2,\dots ,x_n  \quad s.t. \quad ||x_i-x_j||=d_{ij}, (i,j)\in S,  \label{orginal_problem}
\end{eqnarray}
where $||\cdot ||$ denotes a norm in a real vector space and it is calculated as the Euclidean distance 2-norm in this paper. (\ref{orginal_problem}) can be reformulated as a mathematical global optimization problem (GOP)
\begin{eqnarray}
(\mathcal{P} ) \quad &\min P(X)=\sum_{(i,j)\in S} w_{ij} (||x_i-x_j||^2 -d_{ij}^2 )^2  \label{prime_problem}
\end{eqnarray}
in the terms of finding the global minimum of the function $P(X)$, where $w_{ij}, (i,j)\in S$ are positive weights, $X = (x_1, x_2, \dots, x_n)^T \in \mathbb{R}^{n\times 3}$ \cite{more1997} and usually $S$ has many fewer than $n^2/2$ elements due to the error in the theoretical or experimental data \cite{zou1997,grosso2009}. There may even not exist any solution $x_1, x_2, \dots, x_n$ to satisfy the distance constraints in (\ref{orginal_problem}), for example when data for atoms $i, j, k \in S$ violate the triangle inequality; in this case, we may add a perturbation term $-\epsilon^TX$ to $P(X)$:
\begin{eqnarray}
(\mathcal{P}_{\epsilon} ) \quad &\min P_{\epsilon}(X)=\sum_{(i,j)\in S} w_{ij} (||x_i-x_j||^2 -d_{ij}^2 )^2 -\epsilon^TX, \label{prime_approx_problem}
\end{eqnarray}
where $\epsilon \geq 0$. So, the molecular model building problem is a problem to get an global minimal solution of (\ref{prime_approx_problem}). Specially, for the amyloid fibril molecular modeling problem, we find after mutations the hydrogen bonds between $\beta$-strands are still maintained but the vdw contacts become very far. Thus, the $d_{ij}$ in (\ref{prime_approx_problem}) should be the sum of vdw radii of atoms $i$ and $j$.\\

After mutations of 3NHC.pdb, 3NVE/F/G/H.pdb by Swiss-PdbViewer, we find the following least vdw contacts should be maintained respectively for the 3NHC, 3NVE/F/G/H models. 3NHC: A.ALA3.CB-G.ALA4.CB, B.ALA4.CB-H.ALA3.CB (A.ALA3.CB and B.ALA4.CB (two anchors) and G.ALA4.CB and H.ALA3.CB (two sensors), MODEL06-08). 3NVE: A.ALA4.CB-G.ALA3.CB, B.ALA4.CB-G.ALA3.CB (MODEL09), or A.ALA4.CB-G.ALA3.CB, A.ALA2.CB-G.ALA3.CB (MODEL10-11). 3NVF: A.GLY2.CA-H.GLY2.CA, A.ALA4.CB-H.GLY2.CA (MODEL12), A.ALA4.CB-H.ALA2.CB, A.ALA2.CB-H.ALA2.CB, A.ALA2.CB-H.ALA4.CB (MODEL13-14). 3NVG: A.GLY2.CA-H.GLY2.CA, A.GLY2.CA-H.ALA4.CB, A.ALA4.CB-H.GLY2.CA (MODEL15), A.ALA2.CB-H.ALA2.CB, A.ALA2.CB-H.ALA4.CB, A.ALA4.CB-H.ALA2.CB (MODEL16-17). 3NVH: A.ALA4.CB-H.ALA4.CB (MODEL18-19). We look at the former as anchor(s) and its partner as sensor(s). Thus in (\ref{prime_approx_problem}) the variables are 3 or 6 and its dual variables are 1 or 2 or 3. Any optimization solver should be able to accurately solve the prion AGAAAAGA amyloid fibril model building problem. Then the MODEL06-19 will be refined by the SDCG optimization program of Amber 10 and the perfect 3NHC, 3NVE/F/G/H prion AGAAAAGA amyloid fibril 3NHC, 3NVE/F/G/H models were got (Fig.s \ref{MODELS06-08}-\ref{MODELS18-19}).
   
\section{CONCLUSION}
\begin{figure}[h!]
\centerline{
\includegraphics[scale=0.2]{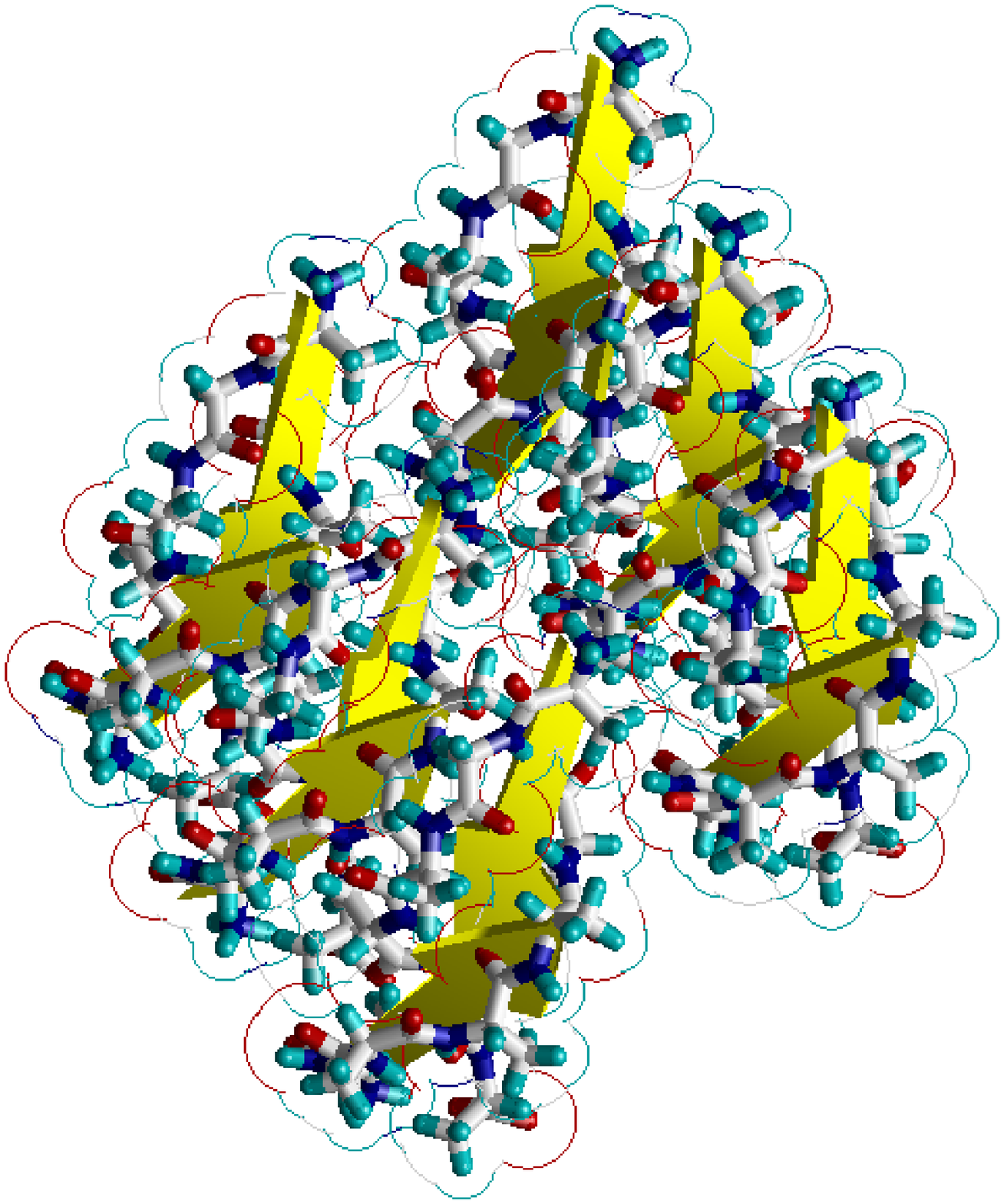}
\includegraphics[scale=0.2]{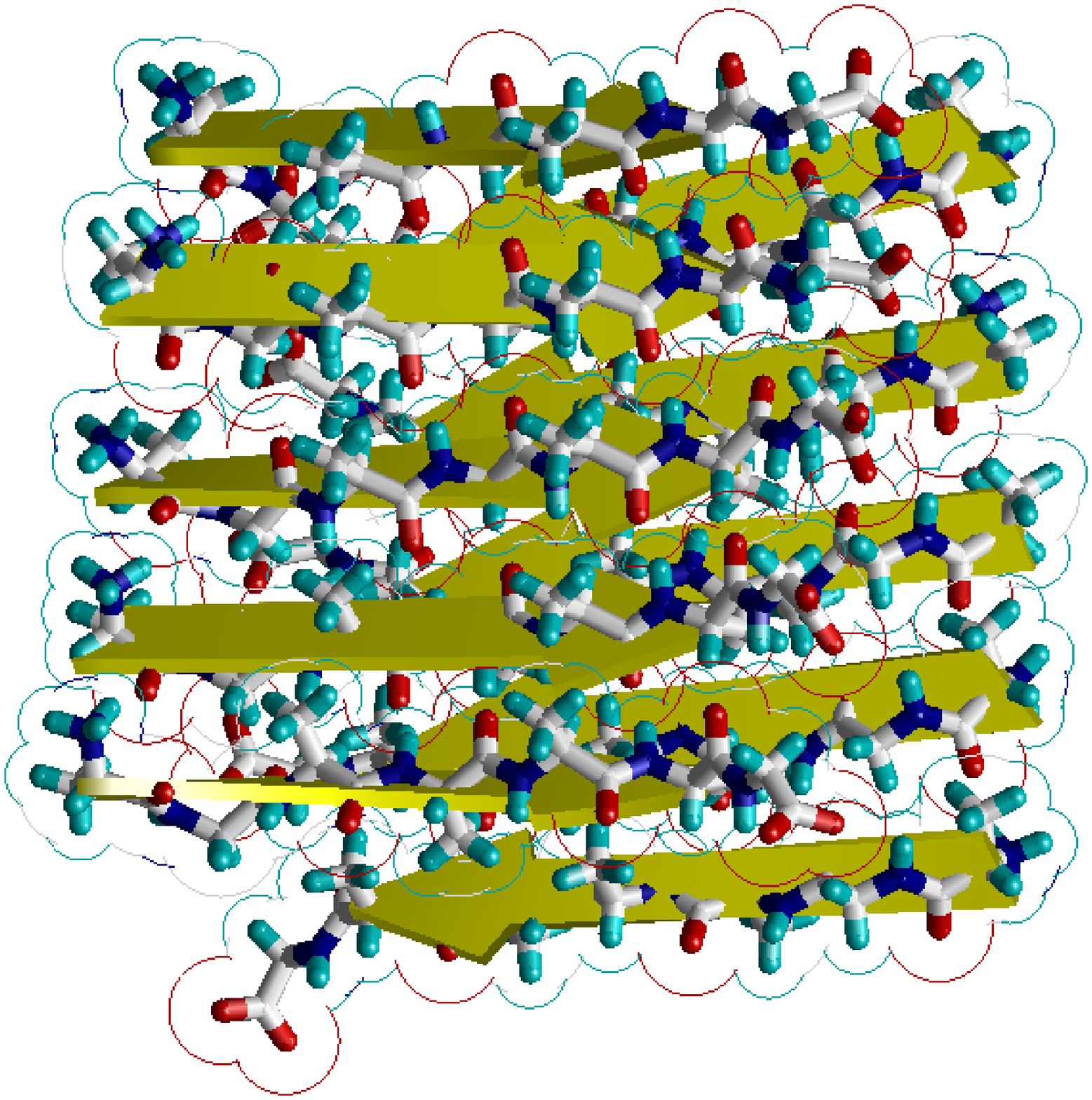}
\includegraphics[scale=0.2]{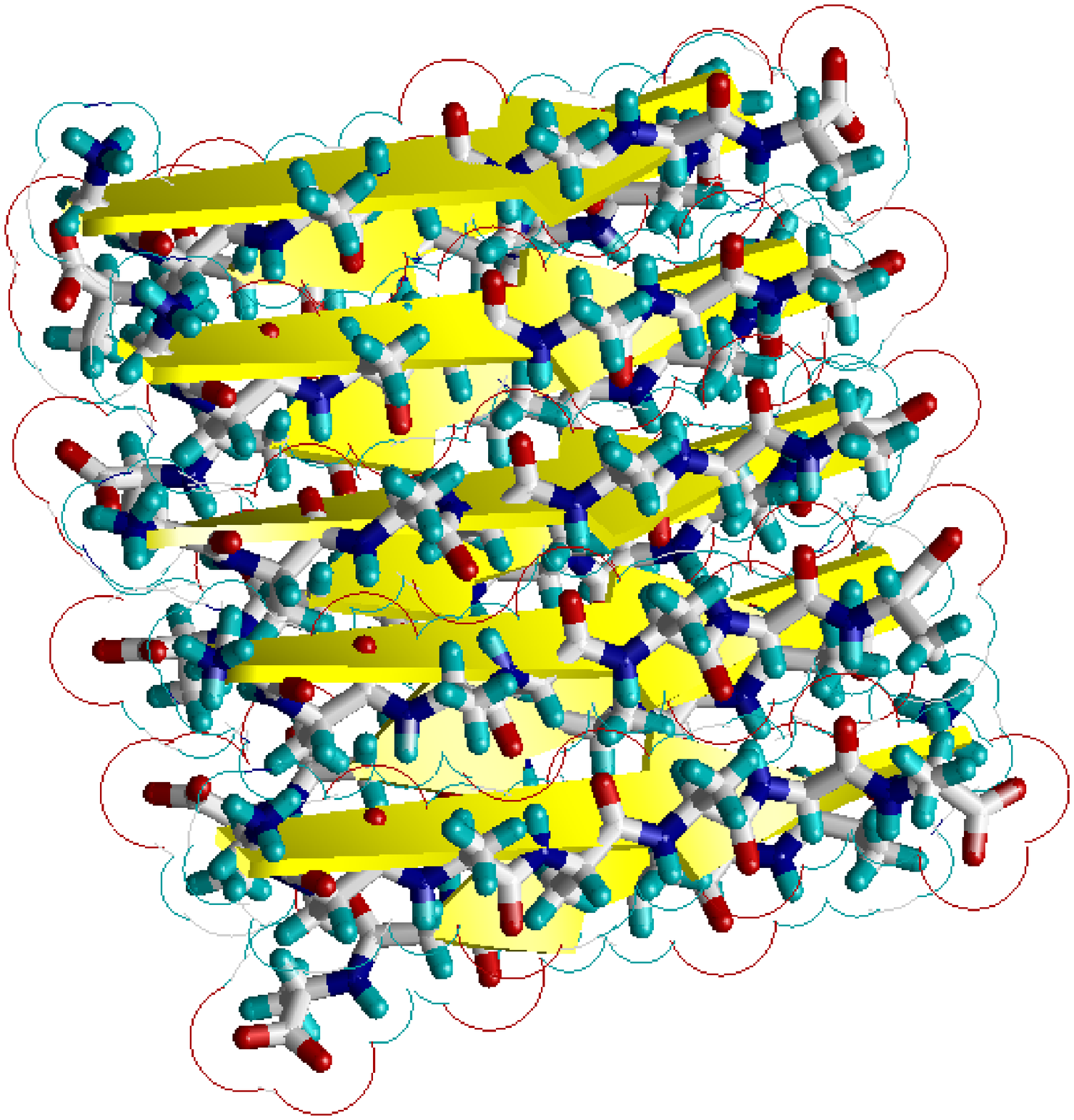}
}
\caption{MODEL01 - MODEL03 for prion (113-120) AGAAAAGA amyloid fibrils.}
\label{MODELS01-03}
\end{figure}

\begin{figure}[h!]
\centerline{
\includegraphics[scale=0.2]{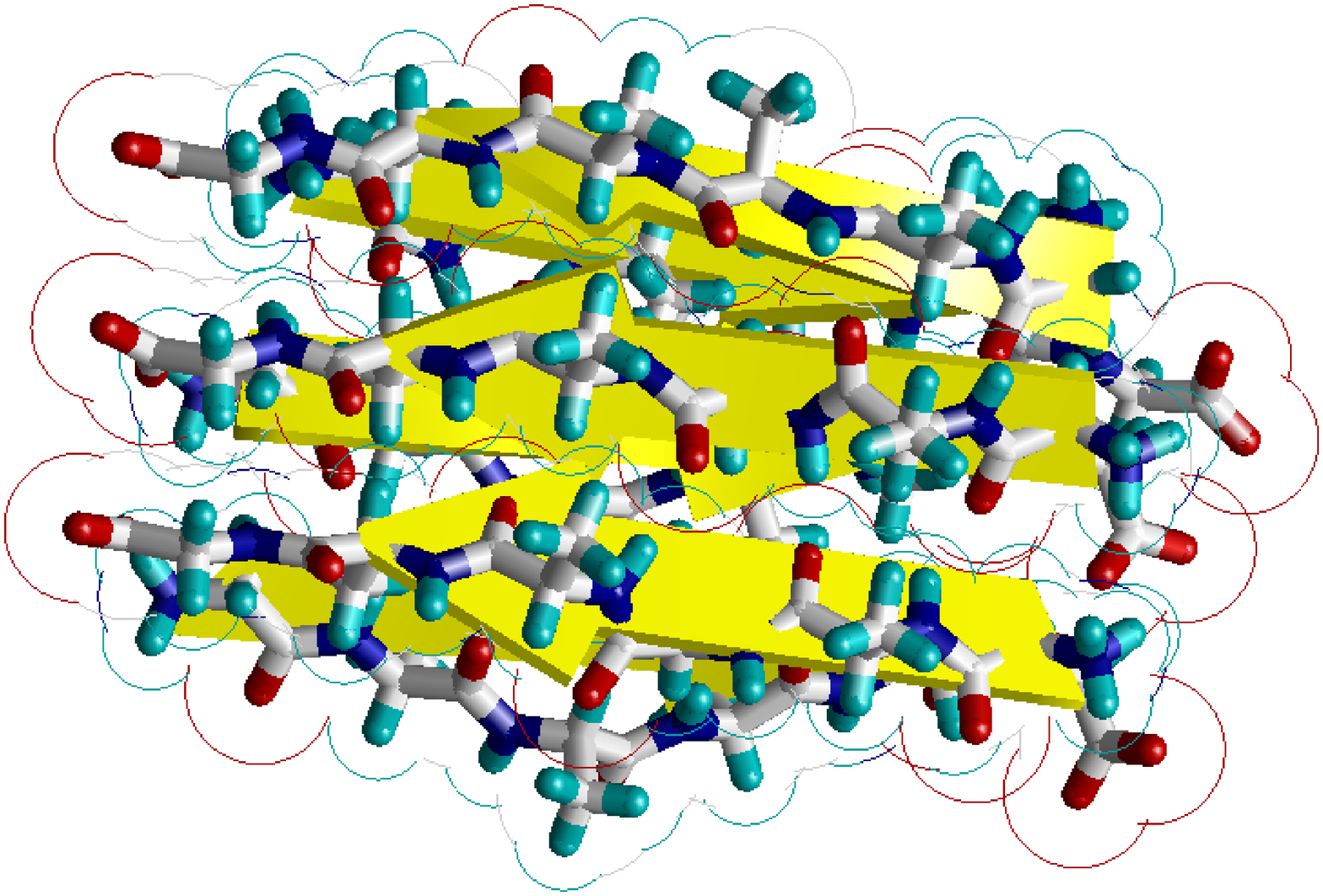}
\includegraphics[scale=0.2]{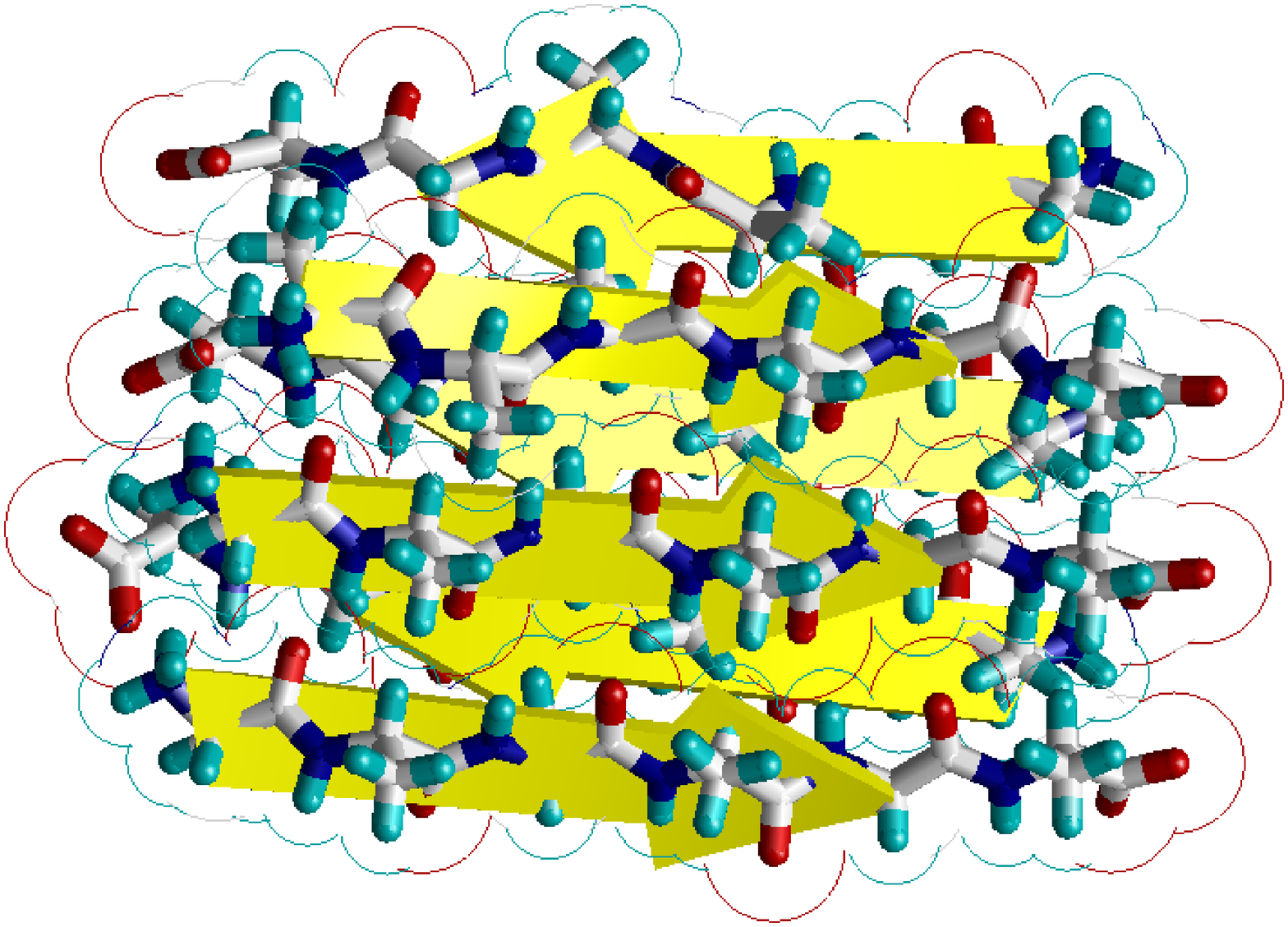}
}
\caption{MODEL04 - MODEL05 for prion (113-120) AGAAAAGA amyloid fibrils.}
\label{MODELS04-05}
\end{figure}

\begin{figure}[h!]
\centerline{
\includegraphics[scale=0.2]{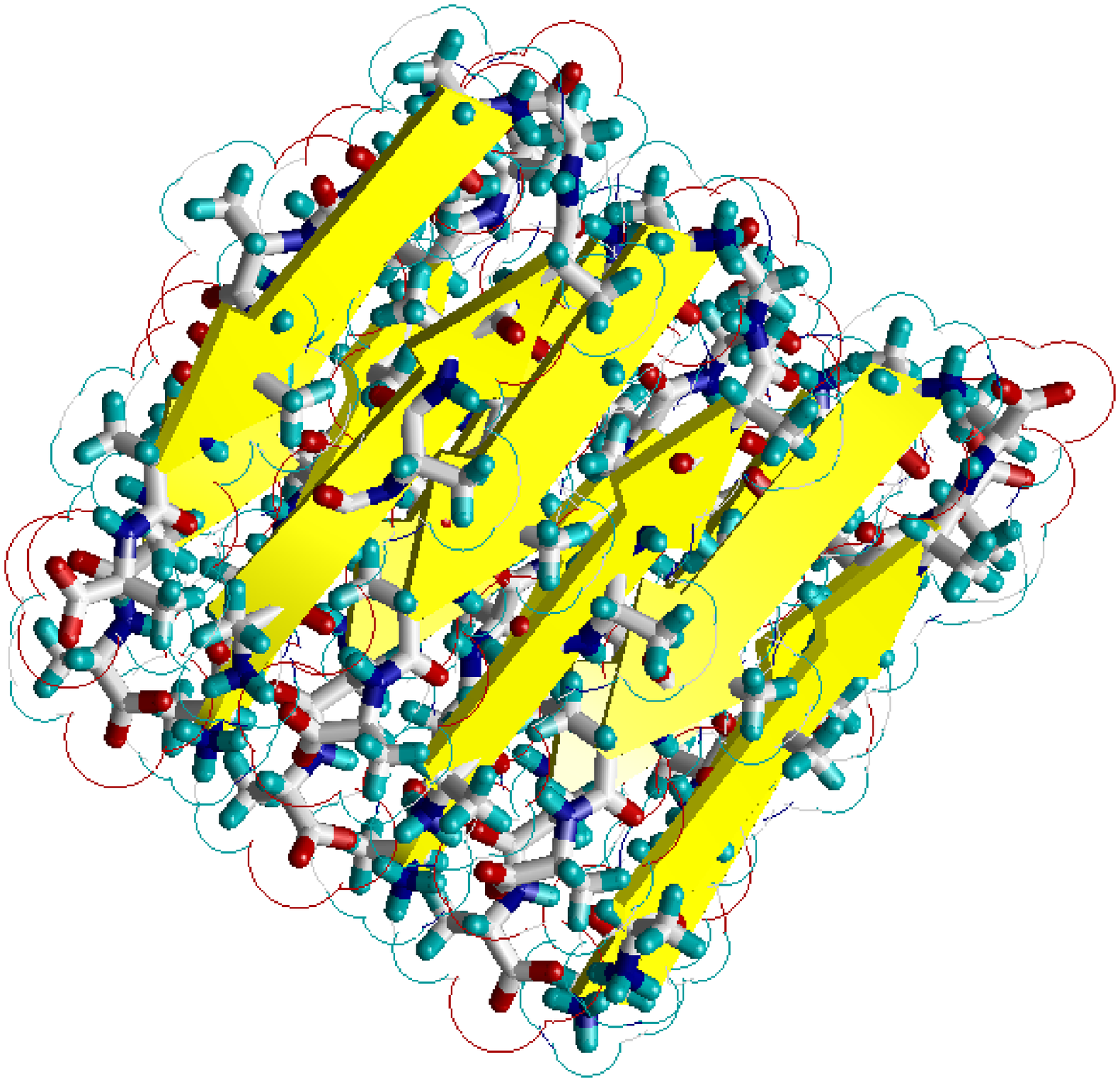}
\includegraphics[scale=0.2]{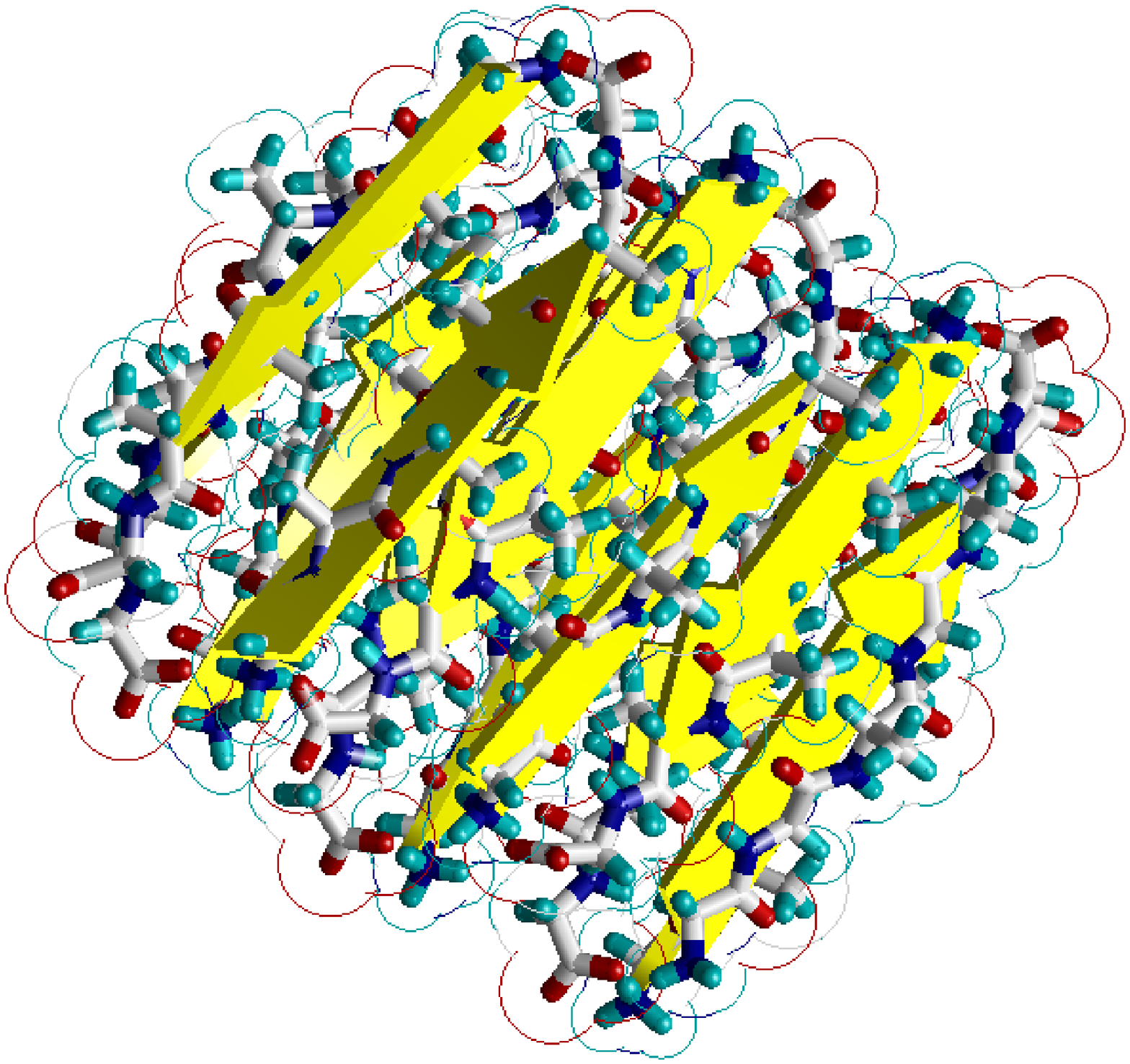}
\includegraphics[scale=0.2]{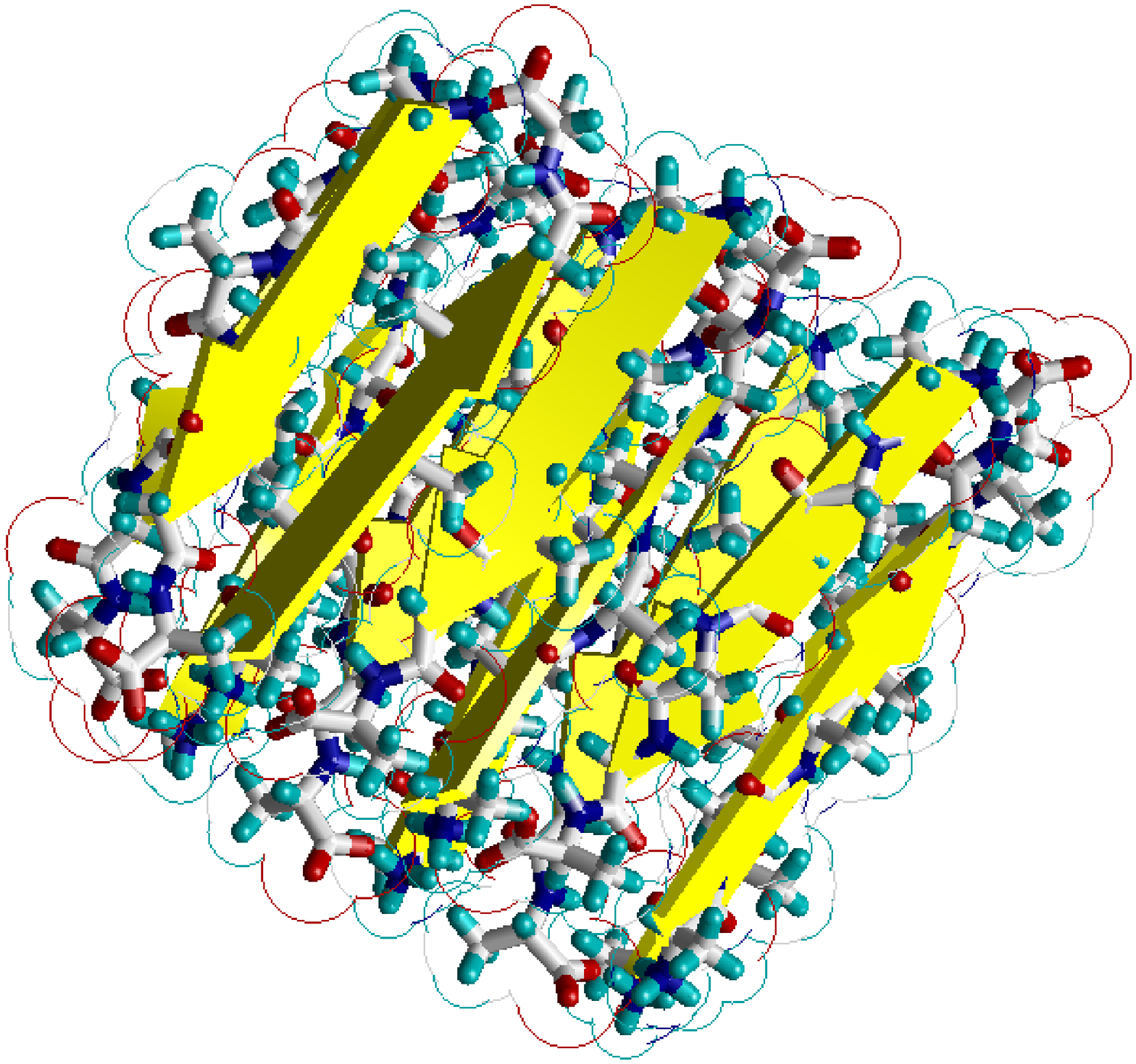}
}
\caption{MODEL06 - MODEL08 for prion (113-120) AGAAAAGA amyloid fibrils.}
\label{MODELS06-08}
\end{figure}

\begin{figure}[h!]
\centerline{
\includegraphics[scale=0.2]{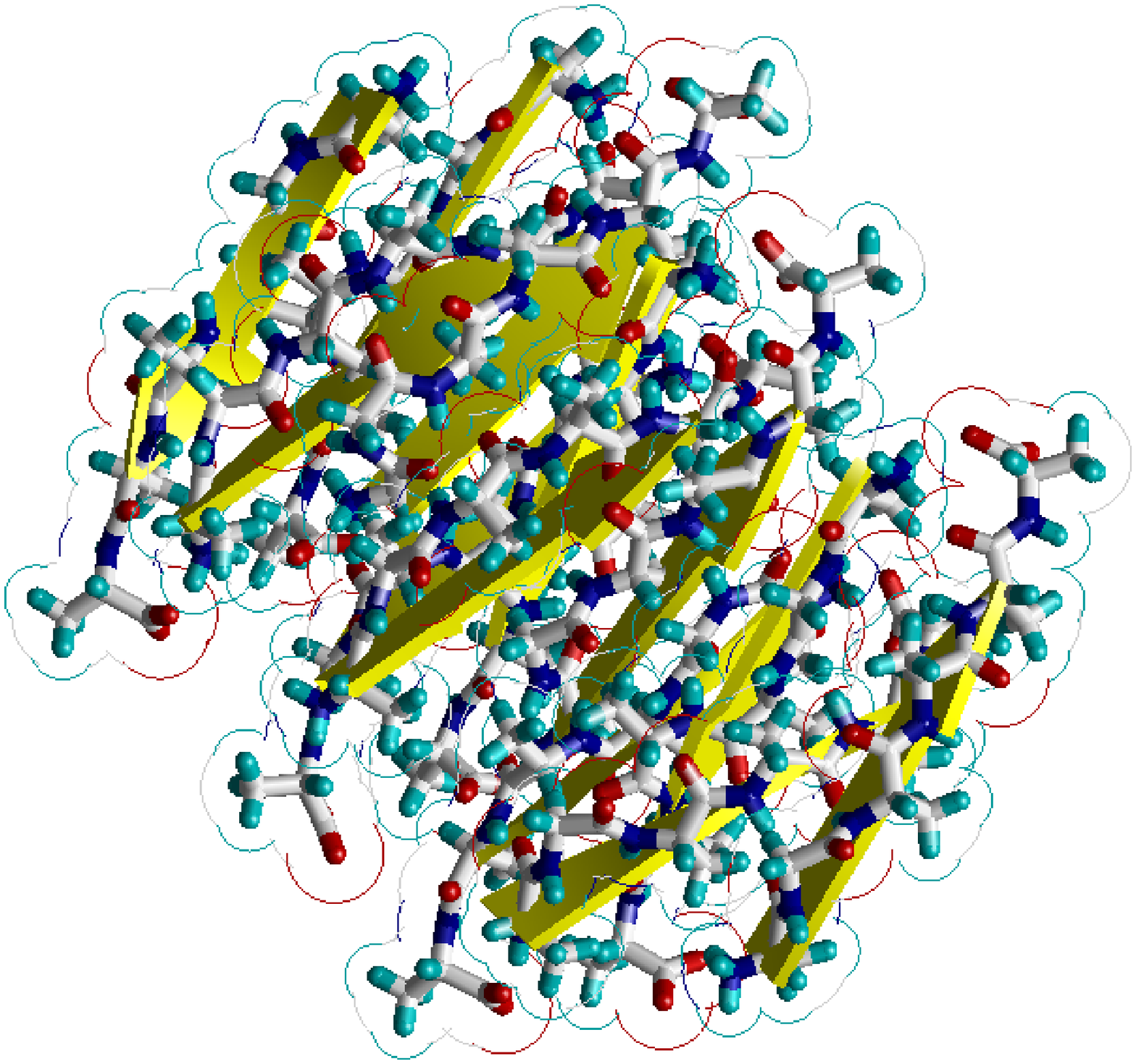}
\includegraphics[scale=0.2]{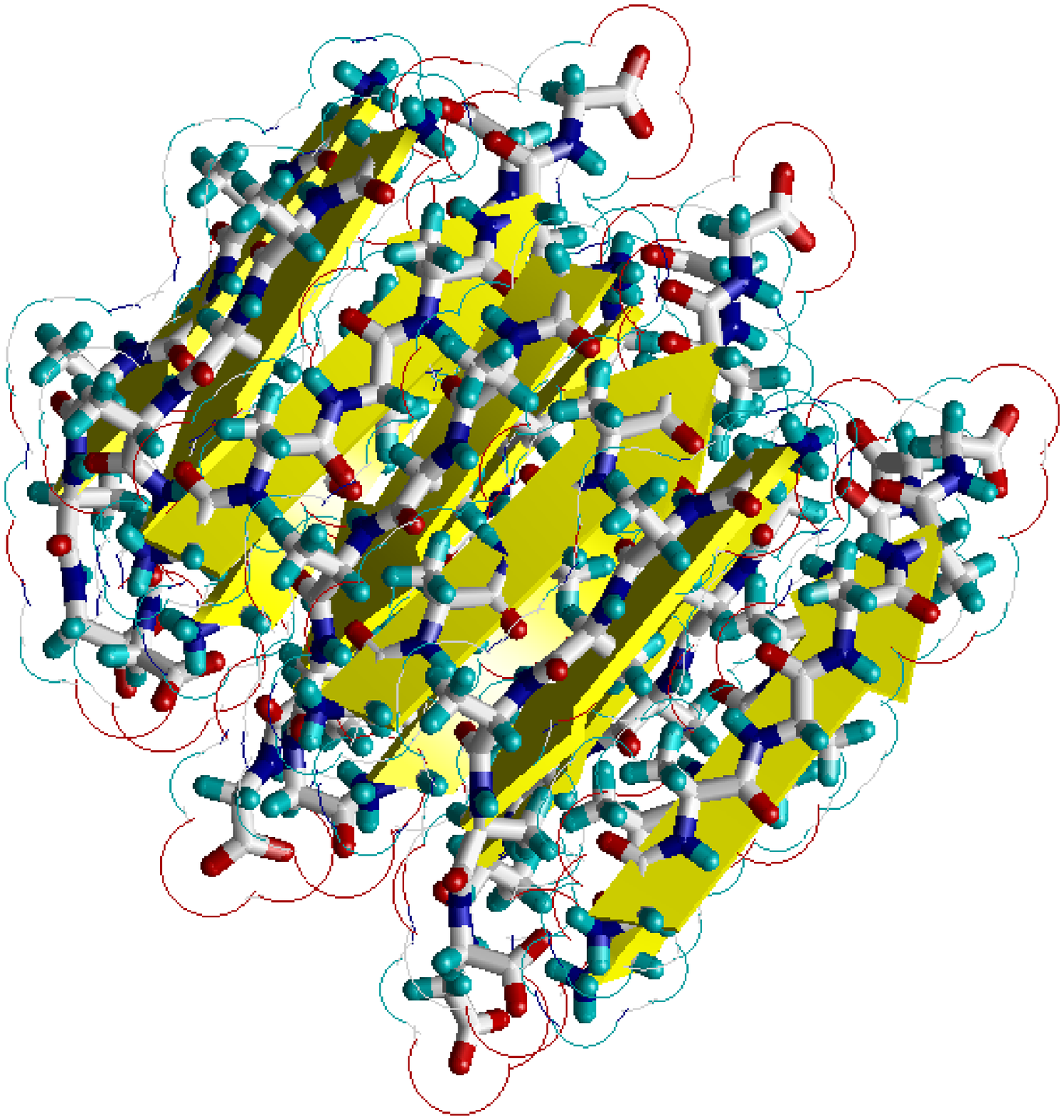}
\includegraphics[scale=0.2]{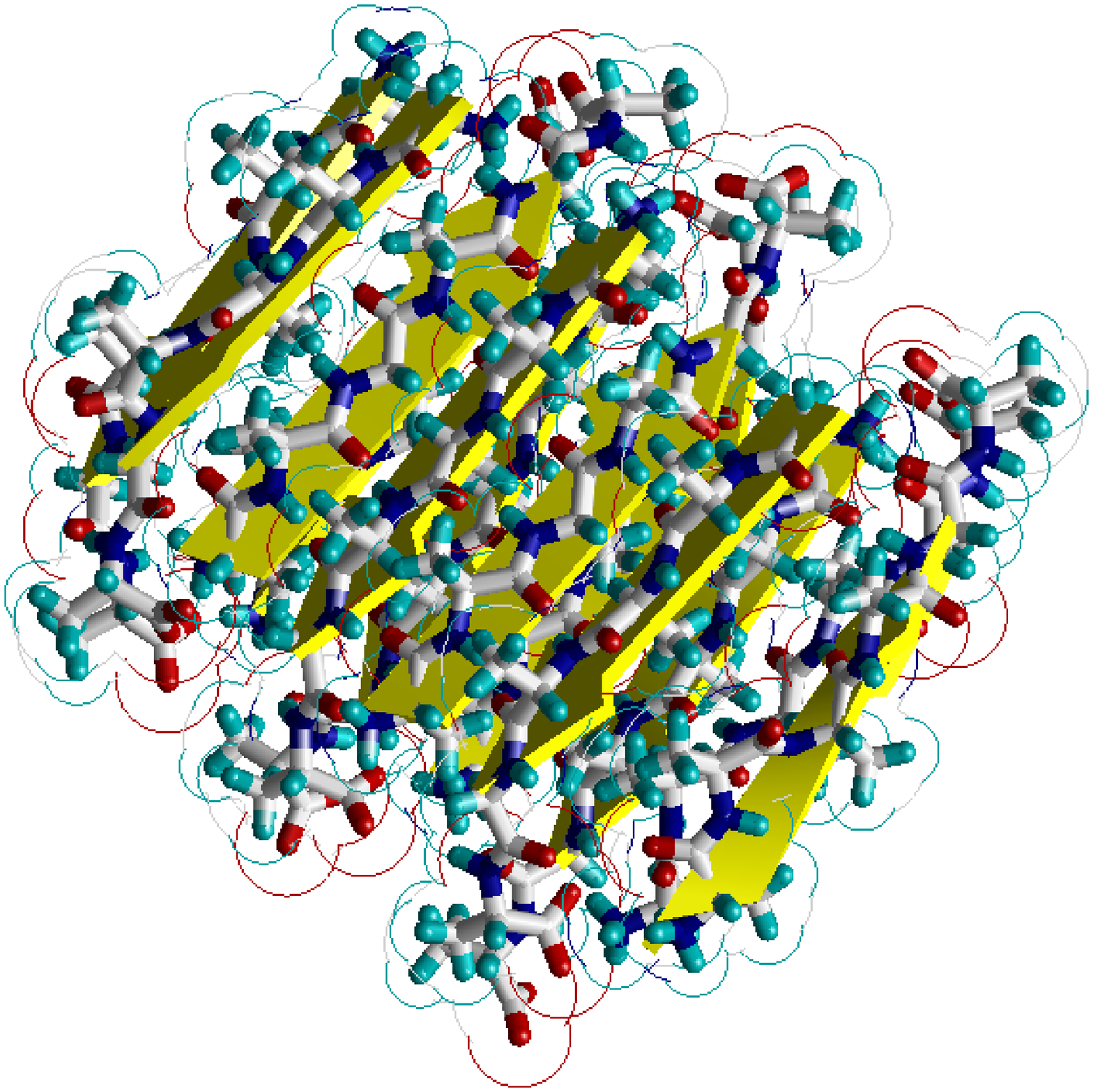}
}
\caption{MODEL09 - MODEL11 for prion (113-120) AGAAAAGA amyloid fibrils.}
\label{MODELS09-11}
\end{figure}

\begin{figure}[h!]
\centerline{
\includegraphics[scale=0.2]{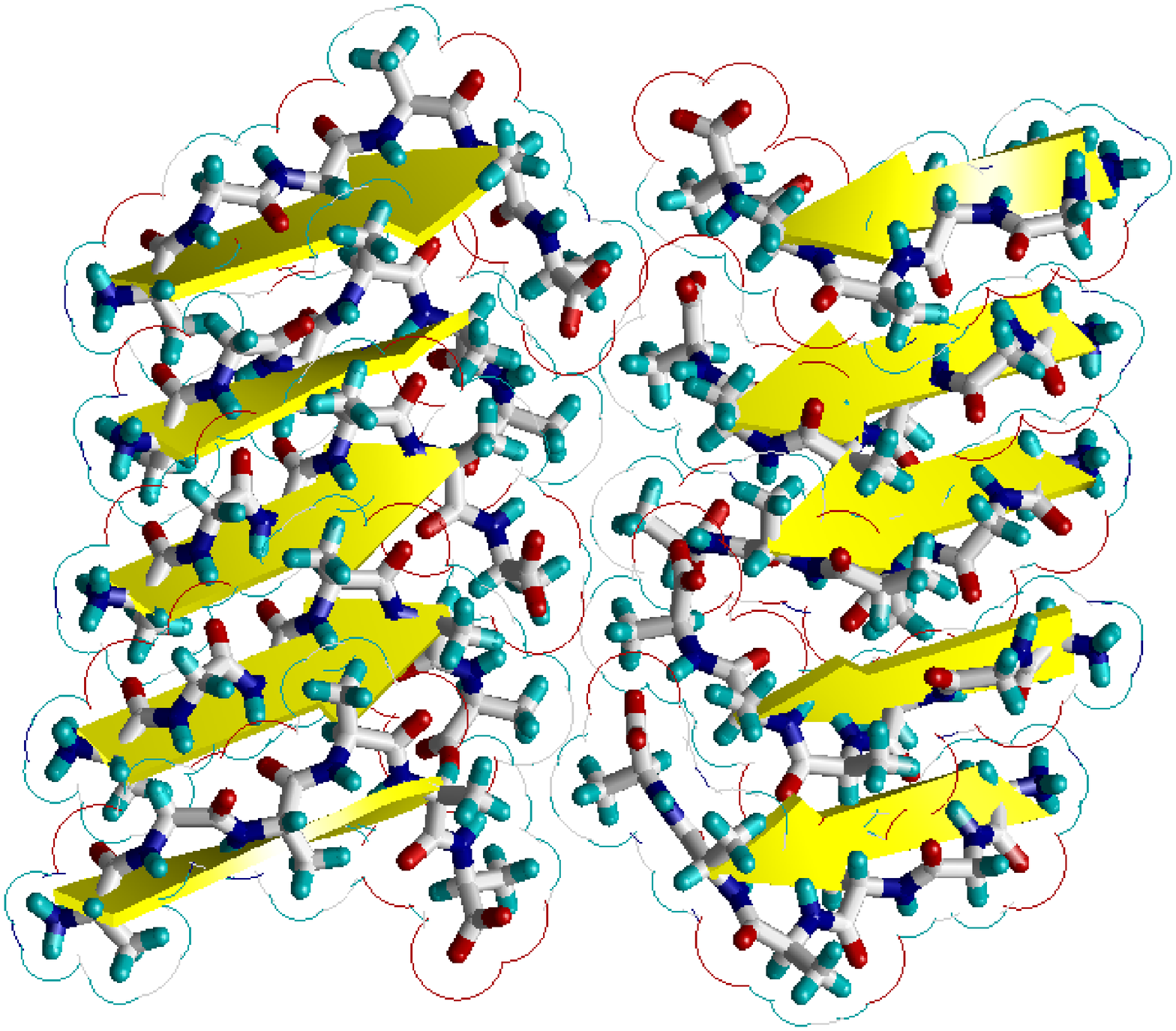}
\includegraphics[scale=0.2]{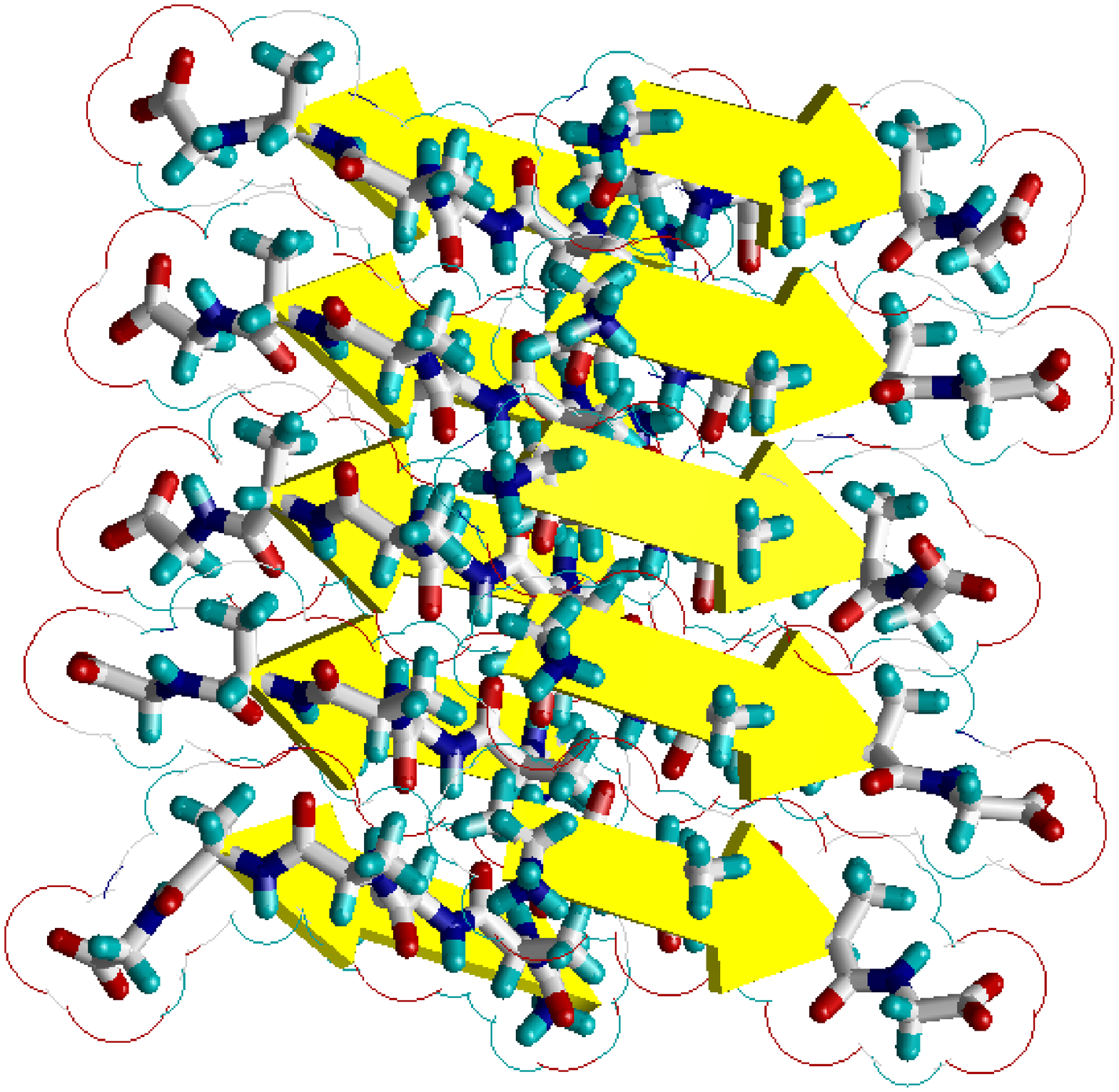}
\includegraphics[scale=0.2]{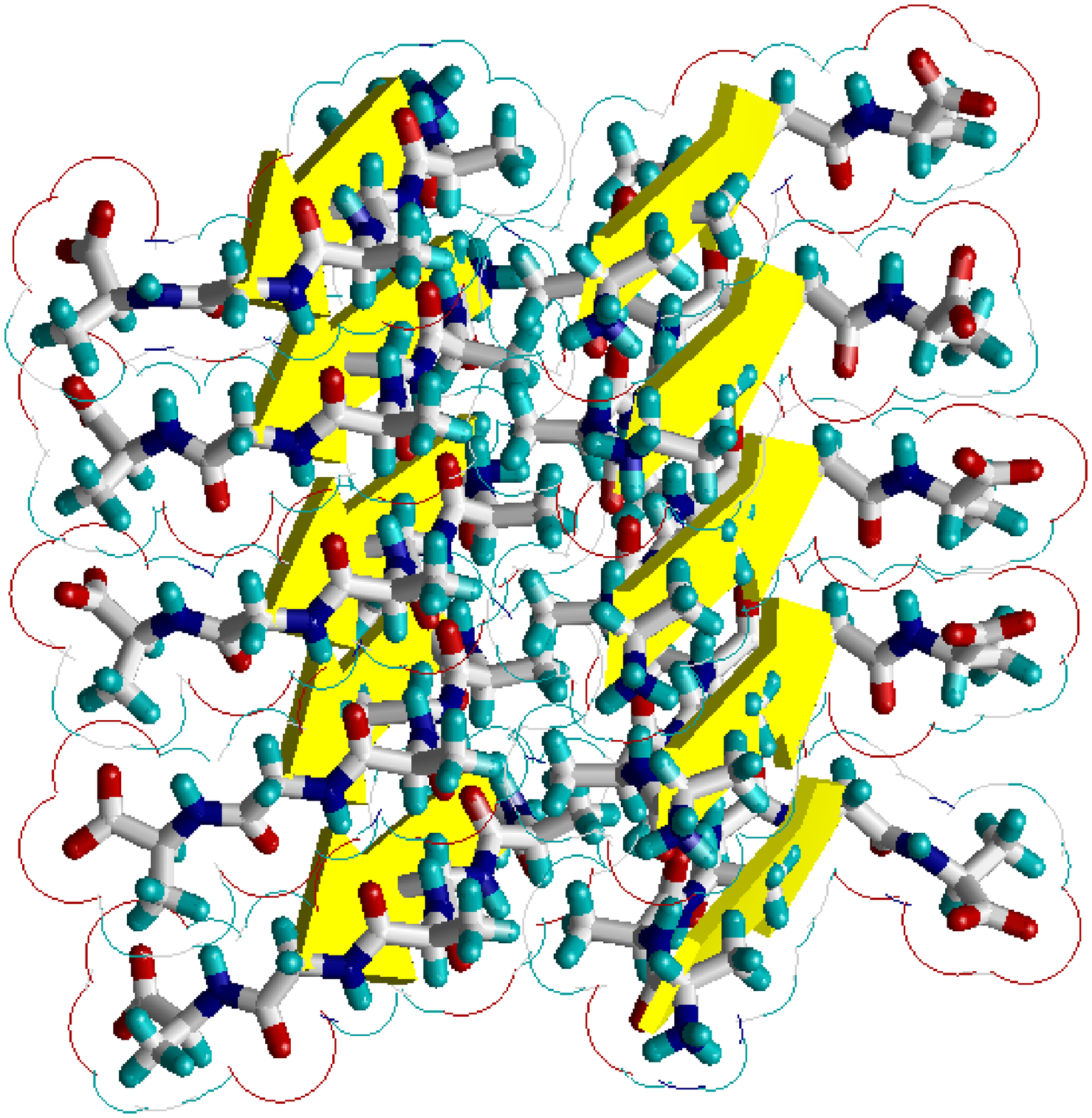}
}
\caption{MODEL12 - MODEL14 for prion (113-120) AGAAAAGA amyloid fibrils.}
\label{MODELS12-14}
\end{figure}

\begin{figure}[h!]
\centerline{
\includegraphics[scale=0.2]{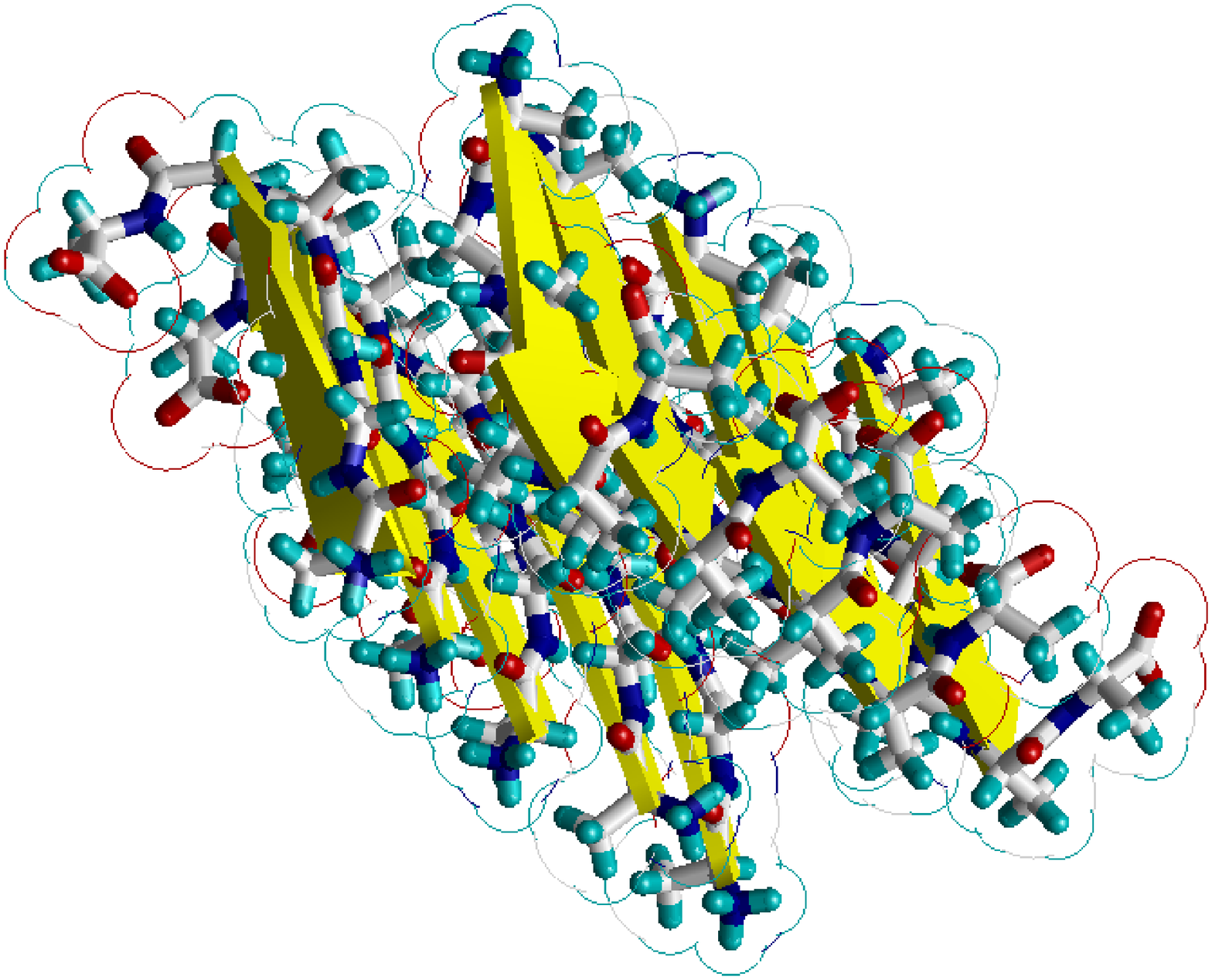}
\includegraphics[scale=0.2]{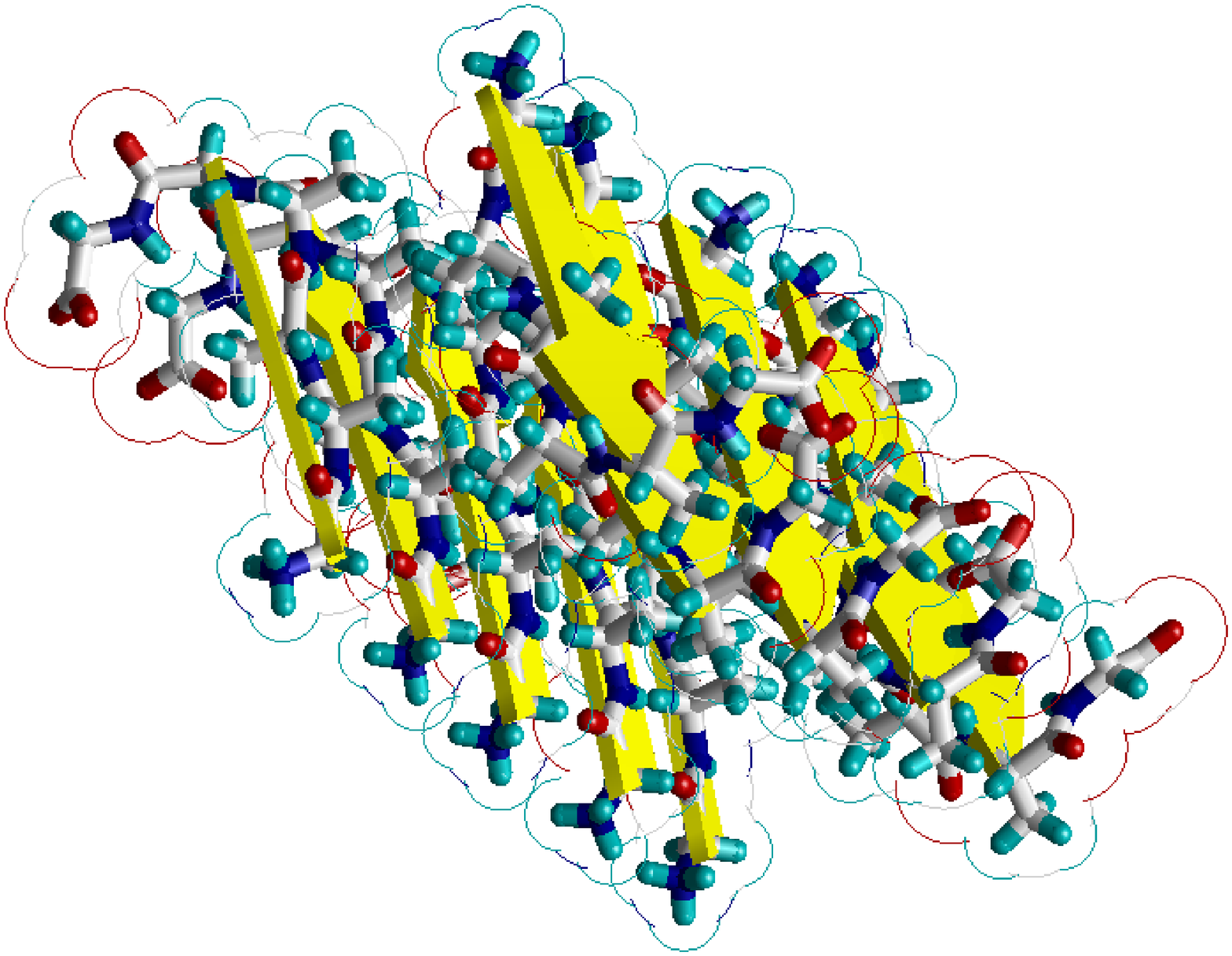}
\includegraphics[scale=0.2]{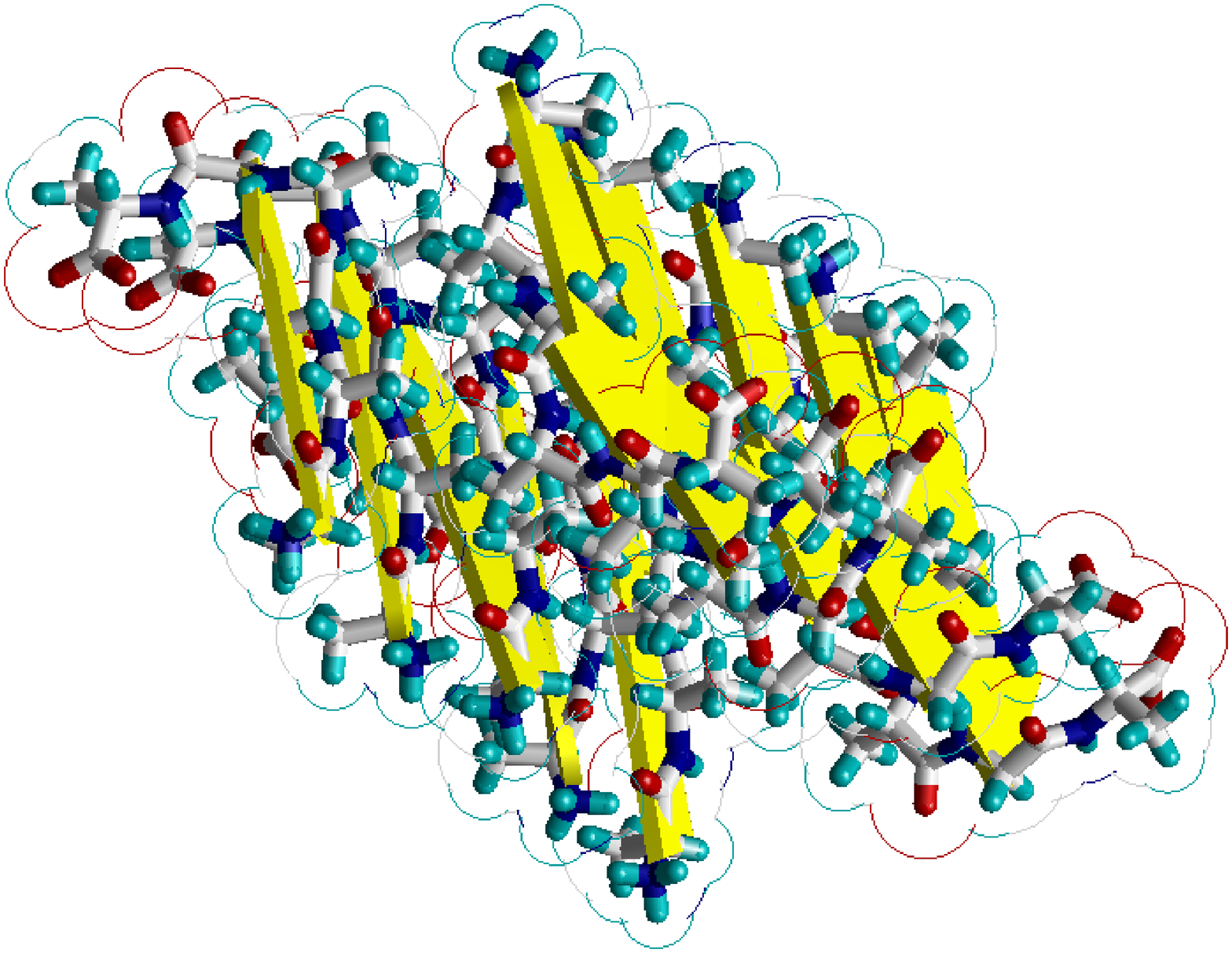}
}
\caption{MODEL15 - MODEL17 for prion (113-120) AGAAAAGA amyloid fibrils.}
\label{MODELS15-17}
\end{figure}

\begin{figure}[h!]
\centerline{
\includegraphics[scale=0.2]{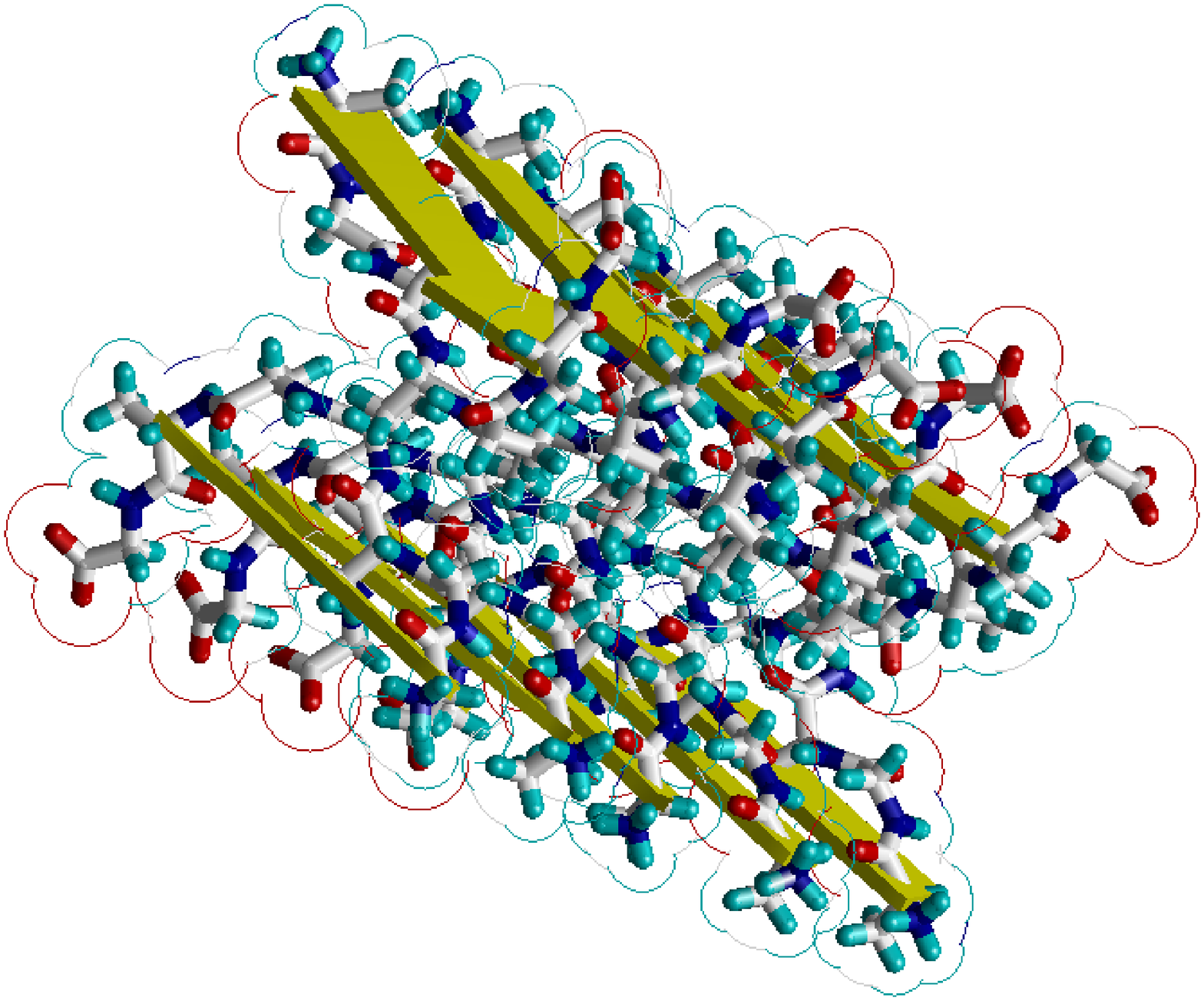}
\includegraphics[scale=0.2]{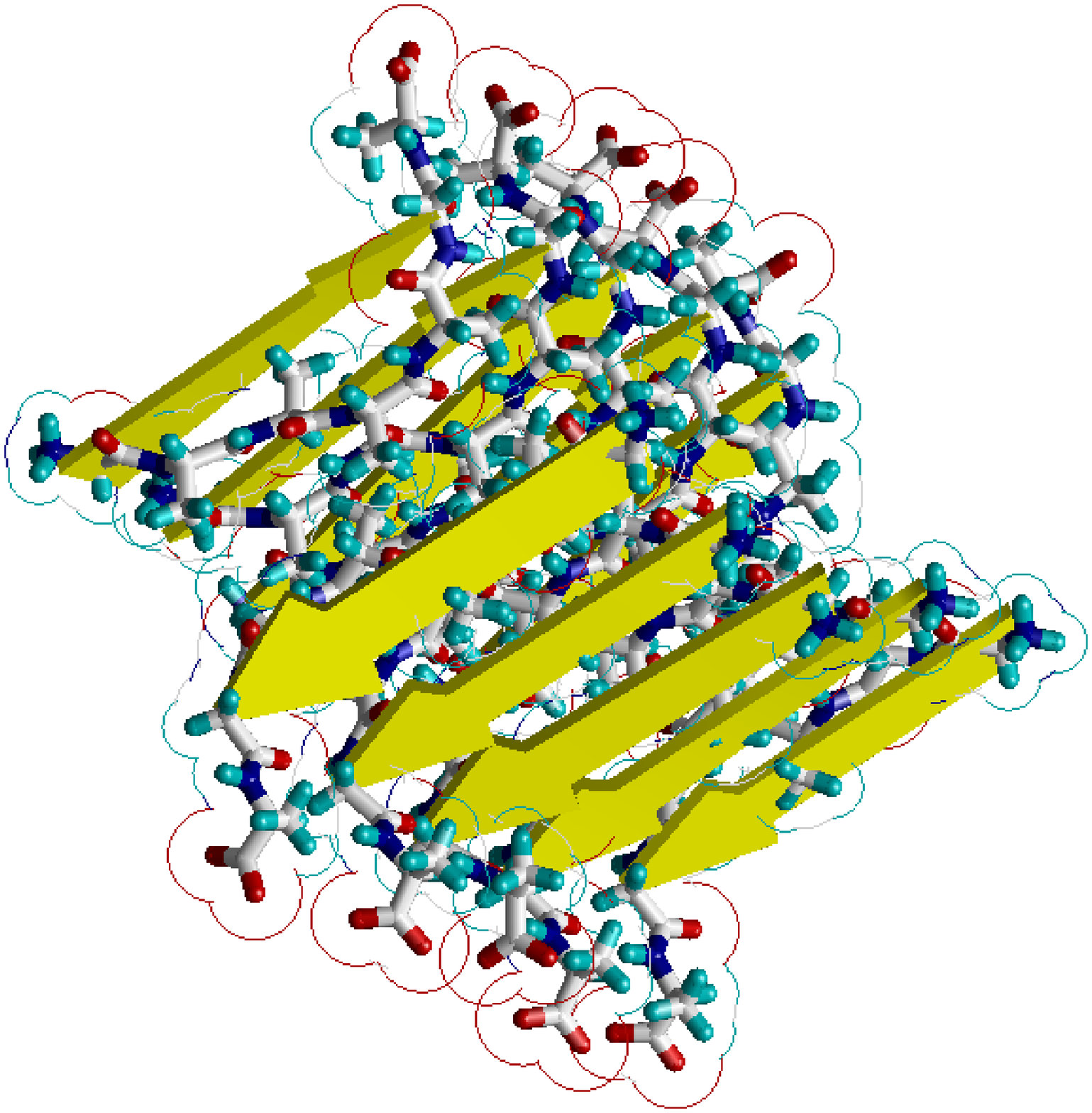}
}
\caption{MODEL18 - MODEL19 for prion (113-120) AGAAAAGA amyloid fibrils.}
\label{MODELS18-19}
\end{figure}

Whenever traditional X-ray crystallography and NMR spectroscopy experimental methods cannot be used to get any structural information of proteins, computational approaches or introducing novel mathematical formulations and physical concepts into molecular biology can significantly stimulate the development of biological and medical science. The numerous optimal atomic-resolution structures of prion AGAAAAGA amyloid fibils reported in Fig.s \ref{MODELS01-03}-\ref{MODELS18-19} have a value to the scientific community in its drive to find treatments for prion diseases or at least be useful for the goals of medicinal chemistry.\\

\noindent {\bf Acknowledgments:} This research was supported by a Victorian Life Sciences Computation Initiative (http://www.vlsci.org.au) grant number VR0063 on its Peak Computing Facility at the University of Melbourne, an initiative of the Victorian Government.

\end{document}